\documentclass[twocolumn,showpacs,preprintnumbers,amsmath,amssymb,floatfix,aps]{revtex4}

\input epsf
\def\vtabstrut{\hbox{\vphantom{$10^{8\strut}$}}}	

\usepackage{graphicx}
\usepackage{dcolumn}
\usepackage{array}
\usepackage{bm}

\begin{document}

\title{Distinguishing cancerous from non-cancerous \\
cells through analysis of electrical noise}

\author{D.C. Lovelady}
\author{T.C. Richmond}
\author{A.N. Maggi}
\author{C.-M. Lo}
\author{D.A. Rabson}
\email[corresponding author: ]{davidra@ewald.cas.usf.edu}
\affiliation{Department of Physics,%
 University of South Florida, Tampa, Florida 33620, USA}

\date{\today}

\begin{abstract}
Since 1984, electric cell-substrate impedance sensing (ECIS) has
been used to monitor cell behavior in tissue culture and has proven
sensitive to cell morphological changes and cell motility. We have taken ECIS
measurements on several cultures of non-cancerous (HOSE) and cancerous
(SKOV) human ovarian surface epithelial cells.  By analyzing the
noise in real and imaginary electrical impedance, we demonstrate
that it is possible to distinguish the two cell types purely from
signatures of their electrical noise.  Our measures include
power-spectral exponents, Hurst and detrended fluctuation analysis,
and estimates of correlation time; principal-component
analysis combines all the measures.  The noise from both cancerous and
non-cancerous cultures shows correlations on many time scales, but
these correlations are stronger for the non-cancerous cells.
\end{abstract}


\maketitle

\section{\label{sec:level1}INTRODUCTION\protect}

Electrical cell-substrate impedance sensing (ECIS) has been in use since
1984 \cite{giaever0} to monitor changes in cell cultures due to spreading
or in response to chemical stimuli, infection, or flow. Applications include
studies of cell migration, barrier function, toxicology, angiogenesis, and
apoptosis. Several papers have noted that impedance fluctuations are associated
with cellular micromotion \cite{luong}. However, we are not aware of any
previous work applying statistical techniques to these fluctuations in order to
distinguish two different cell types. Here, we demonstrate that measures of the
electrical noise from cultures of cancerous and non-cancerous human ovarian
surface epithelial cells distinguish them. We find that the noise in both
cancerous and non-cancerous cultures shows correlations on many time scales,
but by all measures, these correlations are weaker or of shorter range in the
cancerous cultures.

\section{Experimental Methods}
\begin{figure}
\centerline{%
\hbox{\epsfxsize\hsize\epsfbox{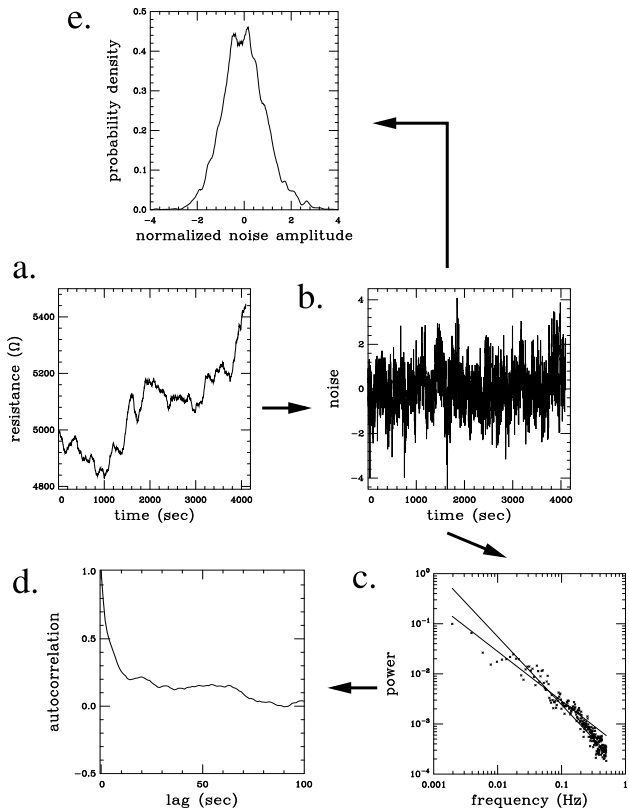}}
}
\caption{\label{fig:overview}
Scheme of data extraction from noise.  (a) Time series of
resistance for one of the experimental runs.  Taking the
discrete time derivative and normalizing to zero mean
and unit variance gives the noise, (b).  The power
spectrum of noise is shown in (c), using overlapping
windows of 256 points in order to reduce scatter.
Fits to the first hundred and last hundred frequencies
estimate low- and high-frequency power-laws, $f^{-\alpha}$.
White noise would have appeared frequency-independent ($\alpha=0$).
The Fourier transform of the power spectrum gives the
autocorrelation, (d), which we fit to a shifted power-law decay
and extract the measure $\beta_0$.  As explained in the text,
subtle differences in the univariate noise distribution (e) (smoothed)
discriminate between cancerous and non-cancerous micromotion.
}
\end{figure}

We used the ECIS system to collect micro-motion time-series data, the 
fluctuations in which are caused by the movements in a confluent layer of live 
cells. The system can be modeled as an RC circuit \cite{giaever1, giaever2, 
lo2, lo1}. The cells are cultured on a small gold electrode
$(5\times 10^{-4}\,{\rm cm}^2)$, which is connected in series to a
1-Megaohm resister, an AC signal generator operating at 1 volt and 4000 Hz, and
finally to a large gold counter-electrode $(0.15\,{\rm cm}^2)$. This network is
connected in parallel to a lock-in amplifier, and the in-phase and out-of-phase
voltages are collected once a second, from which we extract time series of
resistance and capacitive reactance (Figure \ref{fig:overview}a). In ECIS
experiments, the fluctuations in complex
impedance come primarily from
changes in intercellular gaps and in the narrow spaces between the cells and the
small gold electrode \cite{giaever2, lo1, lo2}. A current of about one microamp
is driven through the sample, and the resulting voltage drop of a few millivolts
across the cell layer has no physiological effect: this is a noninvasive, in
vitro-technique. An ovarian cancer line (SKOV3) and a normal human ovarian
surface epithelial (HOSE) cell line (HOSE15) were provided by Dr.\ Samuel Mok
at Harvard Medical School. These cells were grown in M199 and MCDB 105 (1:1)
(Sigma, St. Louis, MO) supplemented with 10\% fetal calf serum (Sigma), 2mM
L-glutamine, 100 units/ml penicillin, and 100 microgram/ml streptomycin under
5\% CO$_2$, and a 37$^\circ\,$C, high-humidity atmosphere. For ECIS micro-motion
measurements, cells were taken from slightly sub-confluent cultures 48 hours
after passage, and a mono-disperse cell suspension was prepared using standard
tissue-culture techniques with trypsin/EDTA. These suspensions were
equilibrated at incubator conditions before addition to the ECIS electrode
wells. Confluent layers were formed 24 hours after inoculation, resulting in a
density of $10^5 {\rm cell}/{\rm cm}^2$.

%

Figure \ref{fig:overview}a shows a representative 4096-second
run (just over one hour) measuring the real part of impedance
as a function of time; the example shows a HOSE culture, but
to the eye, SKOV cultures do not appear very different.  While
the example shows increasing resistance with time, others show
a decrease; at this time scale, there is no evidence for
an overall trend.
We collected, under similar conditions, 18 time series for HOSE cultures,
of which 16 went for 8192 seconds and two for
4096 seconds.  Each 8192-second run was split in two halves, so that
effectively we had thirty-four 4096-second runs; however, where appropriate
in the analysis below, we discard the second
halves of the longer runs in order to
avoid inadvertently introducing correlations.  Similarly, for
SKOV cultures we took data in eight 8192-second runs and
ten 4096-second runs, yielding effectively twenty-six 4096-second runs.
We numerically differentiated the resistance and capacitance time
series to obtain noise time series for each, which we normalized to zero
mean and unit variance (Figure \ref{fig:overview}b).

\section{Statistical Measures of Noise}
We seek information from the normalized noise series.
The first question to pose
is whether the noise can distinguish cancerous from non-cancerous
cultures, but more generally the measures we extract may be used
to test models of cell micromotion.  Broadly,
such models may be characterized by short-term and long-term
correlation, so we look at several measures for each.

First, the power spectral density (Figure \ref{fig:overview}c)
looks very much more like ``pink noise'' than ``white noise;''
that is, it shows signs of long-time correlations.  A log-log
plot of spectral density against frequency, $f$, suggests
an intensity going as $f^{-\alpha}$ in the low-frequency limit.
(We discuss
below the extent to which a true white-noise process may mimic
pink noise due to the finite time of a run.)  For each run, we split
the 4096 noise amplitudes into overlapping bins of 256 seconds,
multiplied by a Hann window, Fourier transformed, and squared,
averaging the resulting spectra in order to reduce scatter
\cite{NR}.

As in the example
of the figure, some runs show a crossover between
low- and high-frequency values for $\alpha$, which we estimated
with least-squares straight-line fits of power at the first 100 (excluding
zero frequency and the very lowest frequency) and last
100 frequencies.
In many runs, low- and high-frequency
alpha estimates were equal, within fitting errors.
Table \ref{tab:power} summarizes
the results, giving in the columns labeled ``ave'' the means over
all HOSE runs or all SKOV runs for the given measures.  The
differences between alphas for HOSE and SKOV, both low- and high-frequency,
exceed several standard errors (or standard deviations of the mean,
std$/\sqrt N$, where $N$ is the number of runs).
Moreover, the Student-$t$-test
and Kolmogorov-Smirnov test show that the HOSE and SKOV
populations differ \footnote{%
Typically, Kolmogorov-Smirnov is taken
to reject
the (null) hypothesis that two populations were drawn from
the same distributions
if it yields a probability less than 5\%.  Three of the four
alpha measures meet this criterion.
As a control test, half of HOSE runs were checked against the
other half
and SKOV against SKOV, and in every case the alpha
measurements were compatible with the null hypothesis, as
expected.}.
The low-frequency exponents are more significant.
The fact that these measures are larger for HOSE than for SKOV
suggests a difference in long-time correlations in micromotion
and is consistent with the hypothesis that non-cancerous
HOSE cells move in a more orderly manner than cancerous SKOV.


A non-zero $\alpha_{\rm low}$ is indicative of long-time, ``fractal''
\cite{west}, correlation, but as Rangarajan and Ding \cite{Rangarajan00}
point out, relying on power-law behavior alone can lead to incorrect
identification of such correlations when none exist. Two related measures are
the Hurst exponent and the exponent of detrended fluctuation analysis
\cite{mandelbrot69,west,feder,peng0,peng1,dfa};
both methods split the time series of noise
into bins of duration $T$, then determine how a measure scales with $T$. For
Hurst, one subtracts the mean from all the data in a bin and characterizes that
bin by its standard deviation, $S$. The series is integrated, and the minimum
value subtracted from the maximum, yielding the range, $R$. For each bin, one
records the ratio $R/S$ and averages over bins of the same size. The procedure
is repeated for successively larger bins ($T$). A straight-line fit to a
log-log plot of $R/S$ against bin size $T$ reveals a power law,
$R/S\sim T^H$, where $H$ is the Hurst exponent. Detrended fluctuation
analysis runs along similar lines, but within each bin one subtracts a best-fit
line, thus detrending the data. The data in the bin are then characterized by
standard deviation $S\sim T^D$,
where $D$ is the DFA exponent. Table \ref{tab:HD}
shows the results; again, with high confidence (based particularly on Student's
$t$-test) we can conclude that HOSE and SKOV noise come from different
distributions. However, since the means are separated by less than a
population standard
deviation, many runs (of 4096 seconds) would be necessary to determine the
provenance of {\it one\/} particular culture.

{
\begin{widetext}
\begin{table*}[t]
\caption{\label{tab:power}%
Power-spectral
measures of HOSE (non-cancerous) and SKOV (cancerous)
resistive and capacitive noise series.  Shown are estimates
for $1/f^\alpha$ behavior at high and low frequencies.
The means of the alphas differ by
many standard errors (std/$\sqrt N$),
allowing us to distinguish the populations composed of $N$ runs,
although not by enough to distinguish reliably a {\it single\/}
HOSE run from a {\it single\/} SKOV.
The $F$-test and $t$-test give the probabilities
that the variances and means of the distributions of values of
$\alpha$
would differ by as much as or more than they do if the two populations
had come from the same Gaussian distribution.
KS gives the probability under the
Kolmogorov-Smirnov test that the two populations' cumulative
distributions could differ as much as they do.
Small probabilities indicate that the
populations differ; a probability of $0.$ means $<10^{-6}$.  $N=34$ for
HOSE, $N=26$ for SKOV.  In all cases, we
apply the approximate $t$-test for distributions with unequal
variances \cite{NR}.
}
\begin{ruledtabular}
\begin{tabular}{lddddddddd}
&\multicolumn{3}{c}{HOSE}&\multicolumn{3}{c}{SKOV}%
&\multicolumn{3}{c}{prob.\ from same distribution}\\
\mbox{measure}%
&\multicolumn{1}{r}{\mbox{ave}}%
&\multicolumn{1}{r}{\mbox{std}}%
&\multicolumn{1}{r}{\mbox{std$/\sqrt N$}}%
&\multicolumn{1}{r}{\mbox{ave}}%
&\multicolumn{1}{r}{\mbox{std}}%
&\multicolumn{1}{r}{\mbox{std$/\sqrt N$}}%
&\multicolumn{1}{c}{\mbox{$F$-test}}%
&\multicolumn{1}{c}{\mbox{$t$-test}}%
&\multicolumn{1}{c}{\mbox{KS-test}}\\
\hline
\multicolumn{10}{c}{\strut resistance} \\
\hline
$\alpha_{\rm low}$ & 0.991 & 0.132 & 0.02 & 0.800 & 0.148 & 0.03 & 0.54 & 4.\times10^{-6} & 4.2\times10^{-4}\vtabstrut\\
$\alpha_{\rm high}$ & 1.58 & 0.558 & 0.10 & 1.09 & 0.648 & 0.13 & 0.42 & 4.\times10^{-3} & 0.024 \\
\hline
\multicolumn{10}{c}{\strut capacitance} \\
\hline
$\alpha_{\rm low}$  & 0.909 & 0.0988 & 0.02 & 0.734 & 0.131 & 0.03 & 0.13 & 0. & 9.\times10^{-6}\vtabstrut\\
$\alpha_{\rm high}$ & 1.133 & 0.446 & 0.08 & 0.980 & 0.357 & 0.07 & 0.25 & 0.15 & 0.37 \\
\end{tabular}
\end{ruledtabular}
\end{table*}
%
\begin{table*}
\caption{\label{tab:HD}%
Additional measures of long-time correlation in the noise time series,
Hurst and detrended-fluctuation exponents.  See Table-\ref{tab:power}
caption for column descriptions.
}
\begin{ruledtabular}
\begin{tabular}{lddddddddd}
&\multicolumn{3}{c}{HOSE}&\multicolumn{3}{c}{SKOV}%
&\multicolumn{3}{c}{prob.\ from same distribution}\\
\mbox{measure}%
&\multicolumn{1}{r}{\mbox{ave}}%
&\multicolumn{1}{r}{\mbox{std}}%
&\multicolumn{1}{r}{\mbox{std$/\sqrt N$}}%
&\multicolumn{1}{r}{\mbox{ave}}%
&\multicolumn{1}{r}{\mbox{std}}%
&\multicolumn{1}{r}{\mbox{std$/\sqrt N$}}%
&\multicolumn{1}{c}{\mbox{$F$-test}}%
&\multicolumn{1}{c}{\mbox{$t$-test}}%
&\multicolumn{1}{c}{\mbox{KS-test}}\\
\hline
\multicolumn{10}{c}{\strut resistance} \\
\hline
Hurst $H$ & 0.770 & 0.0442 & 0.008 & 0.744 & 0.0876 & 0.017 & 3.\times10^{-4} & 0.17 & 0.099\vtabstrut\\
DFA $D$ & 0.854 & 0.0473 & 0.008 & 0.806 & 0.0793 & 0.016 & 0.006 & 9.8\times10^{-3} & 0.057 \\
\hline
\multicolumn{10}{c}{\strut capacitance} \\
\hline
Hurst $H$ & 0.792 & 0.0474 & 0.008 & 0.731 & 0.0886 & 0.017 & 9.\times10^{-4} & 3.1\times10^{-3} & 0.012\vtabstrut\\
DFA $D$ & 0.843 & 0.0479 & 0.008 & 0.788 & 0.0748 & 0.015 & 0.017 & 2.5\times10^{-3} & 3.4\times10^{-3} \\
\end{tabular}
\end{ruledtabular}
\end{table*}
\end{widetext}
%
\begin{widetext}
\begin{table*}[t]
\caption{\label{tab:auto}%
Measures of short-time correlation in the noise time series:
the lag at which normalized autocorrelation (see Figure \ref{fig:overview}d)
falls to $1/e$, the first zero-crossing of autocorrelation, and the
exponent $\beta_0$ from fitting the first few lags with a shifted
power law.  See the Table-\ref{tab:power} caption for the statistical labels.
Of these measures, the $1/e$ crossing (in resistance) and the zero crossing
(in capacitance) have the greatest significance in distinguishing
the populations; $\beta_0$ is significant only in the sense that
the {\it scatter\/} is very much greater for cancerous SKOV than
for non-cancerous HOSE.
}
\begin{ruledtabular}
\begin{tabular}{lddddddddd}
&\multicolumn{3}{c}{HOSE}&\multicolumn{3}{c}{SKOV}%
&\multicolumn{3}{c}{prob.\ from same distribution}\\
\mbox{measure}%
&\multicolumn{1}{r}{\mbox{ave}}%
&\multicolumn{1}{r}{\mbox{std}}%
&\multicolumn{1}{r}{\mbox{std$/\sqrt N$}}%
&\multicolumn{1}{r}{\mbox{ave}}%
&\multicolumn{1}{r}{\mbox{std}}%
&\multicolumn{1}{r}{\mbox{std$/\sqrt N$}}%
&\multicolumn{1}{c}{\mbox{$F$-test}}%
&\multicolumn{1}{c}{\mbox{$t$-test}}%
&\multicolumn{1}{c}{\mbox{KS-test}}\\
\hline
\multicolumn{10}{c}{\strut resistance} \\
\hline
$1/e$ & 6.35 & 1.76 & 0.30 & 4.91 & 2.62 & 0.51 & 0.032 & 0.020 & 9.1\times10^{-3}\vtabstrut\\
zero & 132. & 88.0 & 15 & 111. & 115. & 23. & 0.14 & 0.44 & 0.068 \\
$\beta_0$ & 1.18 & 0.565 & 0.10 & 5.11 & 12.8 & 2.50 & 0. & 0.13 & 0.48 \\
\hline
\multicolumn{10}{c}{\strut capacitance} \\
\hline
$1/e$ & 5.77 & 1.40 & 0.24 & 4.40 & 3.96 & 0.78 & 0. & 0.10 & 6.0\times10^{-5}\vtabstrut\\
zero & 194. & 136. & 23. & 97.5 & 111. & 22. & 0.29 & 3.7\times10^{-3} & 8.6\times10^{-3}\\
$\beta_0$ & 1.17 & 1.16 & 0.20 & 1.93 & 3.35 & 0.66 & 0. & 0.28 & 0.71 \\
\end{tabular}
\end{ruledtabular}
\end{table*}
\end{widetext}
}

While $\alpha_{\rm low}$, $H$, and $D$ were designed to estimate correlations
at diverging time scales, short-time correlation is conveniently
determined from autocorrelation, Figure \ref{fig:overview}d, normalized
to unity at zero lag. The lag of first zero crossing provides one natural
measure of when correlation is lost, but since autocorrelation curves may
sometimes reach very small, yet positive, plateaus before crossing zero, we
also measured the lag at which the autocorrelation first crosses $1/e$. In a
model with only short-time correlation, the $1/e$ time estimates the
exponential decay time. However, as we discuss below, we observed significant
deviations from exponential decay, finding better fits to a shifted power-law
decay,
\begin{equation}\label{eqn:beta0}
{\rm autocorrelation}
= \left({{t+t_1}\over{t_1}}\right)^{-\beta_0}\rlap{\quad.}
\end{equation}
We fit autocorrelation,
for lags in the heuristic interval $t=1$ to $t=20$ seconds,
using Levenberg-Marquardt least-squares minimization to this form
to find $\beta_0$.  Table \ref{tab:auto} summarizes results for
the two crossings and $\beta_0$; the last distinguishes the
populations of HOSE and SKOV runs only in that (cancerous) SKOV
shows much greater scatter in $\beta_0$, as measured by the $F$-test.
Both crossings vary greatly from run to run, but the $1/e$ crossing
in resistance and zero crossing in capacitance distinguish
the populations of HOSE and SKOV experiments at better than the 95\%
confidence level as measured by Student's $t$-test and the
Kolmogorov-Smirnov test.  In particular, the averaged measures
show shorter crossing times and steeper descents ($\beta_0$)
for SKOV than for HOSE, again consistent with the hypothesis
that the micromotion of cancerous cultures is less correlated
than that of non-cancerous cultures.

\begin{figure}[b]
\centerline{%
\hbox{\epsfxsize\hsize\epsfbox{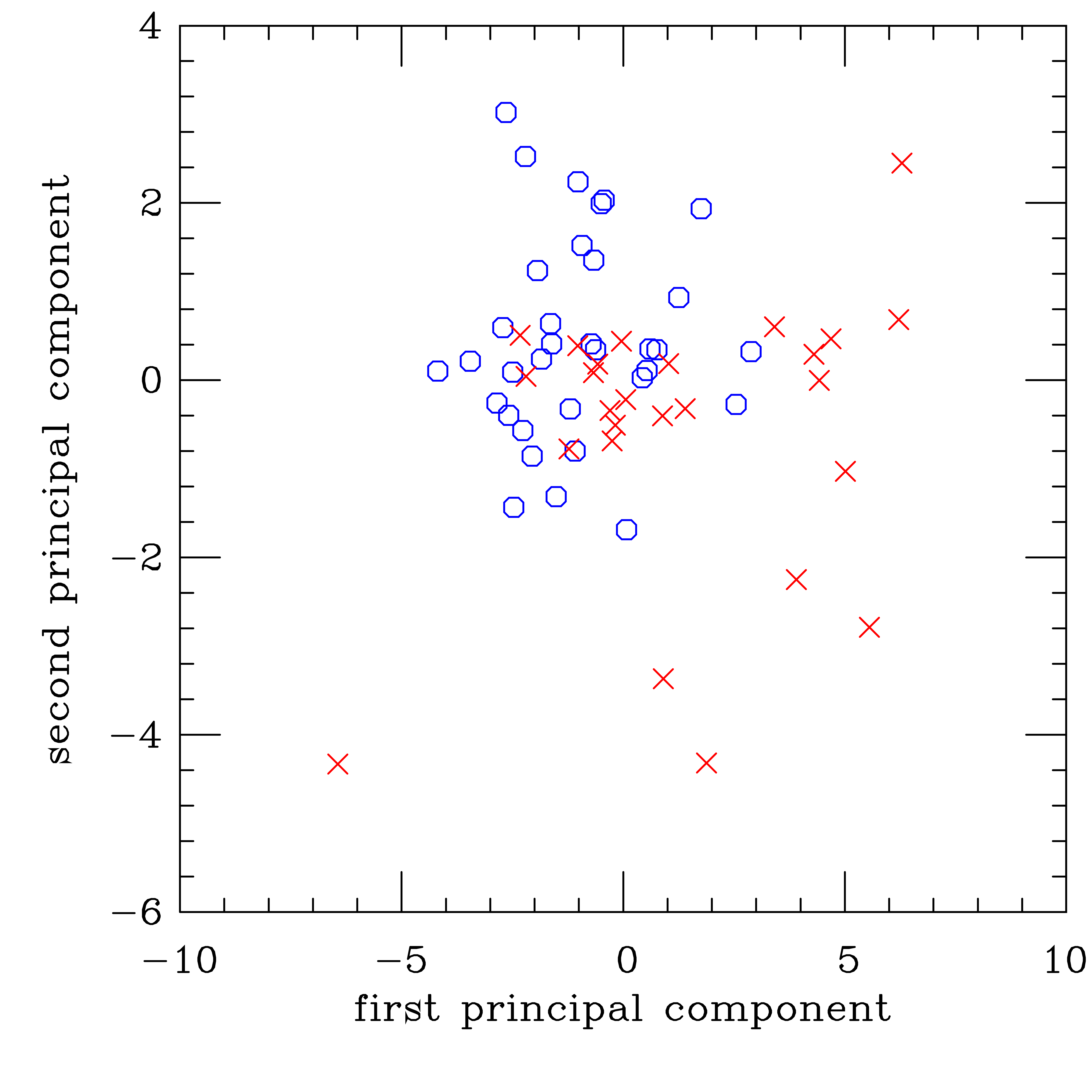}}
}
\caption{\label{fig:pca}
(Color on-line.)  Projection along the first two principal components of the
fourteen-dimensional space determined by Tables \ref{tab:power}--\ref{tab:auto}.
Blue open symbols mark the 34 HOSE runs, red crosses the 26 SKOV
experiments.  As populations, these two sets are distinct, but
the overlap of clusters makes it difficult to distinguish individual
runs in this type of projection.
}
\end{figure}

With the fourteen measures summarized in Tables \ref{tab:power}--\ref{tab:auto},
each run of 4096 seconds can be thought of as a point in a fourteen-dimensional
space. In such problems, the populations might separate into two distinct,
compact clusters \cite[\S4.2]{lewicki};
while the identification of clusters in high-dimensional
spaces remains an open problem in statistical research,
it is common to use the variance-maximizing
principal-component analysis
introduced by Hotelling
to project onto
optimal subspaces, usually taken to be two-dimensional \cite{hotelling}. Figure
\ref{fig:pca} plots the first two principal components. While the plot shows a
clear difference between the two populations consisting of all runs of HOSE and
all runs of SKOV, overlap between the two clusters makes it difficult to apply
the technique
diagnostically.  We found this problem to be generic:
an exhaustive examination of pairs of principal components (beyond
the first two) produced similar plots, with the two populations
usually less distinct in higher-order components, while
adding or subtracting several measures to the list of fourteen
measures did not improve clustering.

Thus far, the noise measures considered have shown that
electrical noise from HOSE and SKOV experiments have, on average,
different correlations, but they do not provide a reliable
way to determine whether the cells in a single run
of 4096 seconds are HOSE or SKOV.  However, from the normalized
(zero-mean, unit-variance) noise time series of Figure \ref{fig:overview}b,
we can extract a probability distribution of noise amplitudes,
as in Figure \ref{fig:overview}e.  Not surprisingly, the
distribution is approximately Gaussian; however, subtle deviations
from normal form do distinguish HOSE from SKOV, even in a single run,
if we apply the Kolmogorov-Smirnov test directly to the noise.  This
test looks only at distributions of noise amplitudes, rather than
correlations.

To this end, we concatenate the first nine 4096-second HOSE resistance
runs (discarding, for this purpose, the second halves of the 8192-second runs)
to create a HOSE resistance reference distribution.  Similarly, we create a
SKOV resistance reference by concatenating the first nine 4096-second
SKOV runs.
Each of the remaining runs
is tested against the two resistance reference sets.
The same procedure is applied with capacitance data.
In many cases, Kolmogorov-Smirnov does not show a match with
either distribution with high probability, but we can
compare the two probabilities: one typical HOSE run matches
the HOSE reference with probability 0.02 and SKOV with probability
$4.7\times10^{-8}$, so we (correctly) identify this run as HOSE based on
the ratio of probabilities.
Of 56 tested data sets (none of which went into the construction
of the reference sets), 42 (75\%) matched
the correct reference set by this criterion,
an outcome that would happen by chance
with probability approximately $1.2\times10^{-4}$.  We repeated
the procedure using a second collection of four reference sets
(HOSE/SKOV, resistance/capacitance) constructed from
nine runs not used in making the first reference sets.  Of 64 trials
(none used in the new reference sets), 53 (83\%) were identified
correctly, with corresponding
probability $5\times10^{-8}$.  The results from the two sets
of trials are added and summarized in Table \ref{tab:KS}.
We can reduce percentages of incorrect identifications by
insisting on agreement between resistance and capacitance
time series; this lowers the overall incorrect identification rate to 11.7\%,
with a correct rate of 70.0\% and a ``not-sure'' rate of 18.3\%.

Deviations from normality in the noise-amplitude distributions
are characterized in part by kurtosis \cite{Keeping},
which is larger for HOSE than for
SKOV: see Table \ref{tab:kurtosis}.
A possible explanation is that kurtosis here is a proxy
for correlation time: since convergence under the central-limit theorem 
is non-uniform, with a distribution approaching a Gaussian slowly in
the tails as the number of samples increases,
a smaller population will tend to have a larger kurtosis
than a larger one.  All of our runs have the same number of time
steps, but as we have seen, SKOV correlation times are shorter than
HOSE correlation times; thus, a SKOV run could be said to have more
{\it independent\/} time steps than a HOSE run of the same length
and might be expected then to have a smaller kurtosis.
However, this cannot explain the whole effect: as we argue
in Appendix \ref{app:errors}, both kurtosis and Kolmogorov-Smirnov
appear to be better discriminants than a direct measure, the $1/e$ crossing.
This suggests that the univariate noise distribution is more than
just a proxy for correlation time.

{
\begin{table}
\caption{\label{tab:KS}%
Percentages of correct identifications.
The Kolmogorov-Smirnov test is applied to distributions of noise
amplitudes against HOSE and SKOV reference sets.
Two non-overlapping choices of reference sets are used; in
neither case did any trial run figure in a reference set against
which it was tested.
The ``average'' column gives percentages weighted by numbers
of trials (16 HOSE and 12 SKOV in the first set, 18 HOSE
and 14 SKOV in the second).
}
\begin{ruledtabular}
\begin{tabular}{lddd}
&\multicolumn{1}{c}{\mbox{first set}}%
&\multicolumn{1}{c}{\mbox{second set}}%
&\multicolumn{1}{r}{\mbox{average}}\\
\hline
HOSE capacitance & 62.5\rlap{\%} & 72.2\rlap{\%} & 67.6\rlap{\%} \\
HOSE resistance & 87.5 & 66.7 & 76.5 \\
SKOV capacitance & 83.3 & 100. & 92.3 \\
SKOV resistance & 66.7 & 100. & 84.6 \\
\noalign{\vphantom{m}}
all resistance & 78.6 & 81.3 & 80.0 \\
all capacitance & 71.4 & 84.4 & 78.3 \\
\noalign{\vphantom{m}}
all HOSE & 75.0 & 69.4 & 72.1 \\
all SKOV & 75.0 & 100. & 88.5 \\
all & 75.0 & 82.8 & 79.2 \\
\end{tabular}
\end{ruledtabular}
\end{table}
}

{
\begin{table}
\caption{\label{tab:kurtosis}%
Kurtosis averaged over all runs, standard deviation of kurtoses,
and standard deviation of the means.
}
\begin{ruledtabular}
\begin{tabular}{lddd}
&\multicolumn{1}{c}{average}
&\multicolumn{1}{c}{std.}
&\multicolumn{1}{c}{std./$\sqrt N$}\\
\hline
HOSE resistance & 74.4 & 173.1 & 29.7 \\
SKOV resistance & 3.00 & 9.57 & 1.9 \\
\noalign{\vphantom{m}}
HOSE capacitance & 17.6 & 32.2 & 5.5 \\
SKOV capacitance & 0.94 & 1.80 & 0.35 \\
\end{tabular}
\end{ruledtabular}
\end{table}
}

\section{TWO SIMPLE MODELS}
Having motivated and interpreted our measures of noise in terms
of short- and long-time correlations, we now compare our data to
the simplest possible discrete-time models, the binary random walk
with persistence \cite{Fuerth1920}, displaying only short-time correlation, and
a discrete fractional Brownian motion \cite{Mandelbrot68,Rangarajan00}, which
has correlations on all time scales.  For present purposes,
it suffices to consider only the increments rather than the walks themselves;
that is, we compare to Figure \ref{fig:overview}b, not
Figure \ref{fig:overview}a.

\begin{figure}
\centerline{%
\hbox{\epsfxsize\hsize\epsfbox{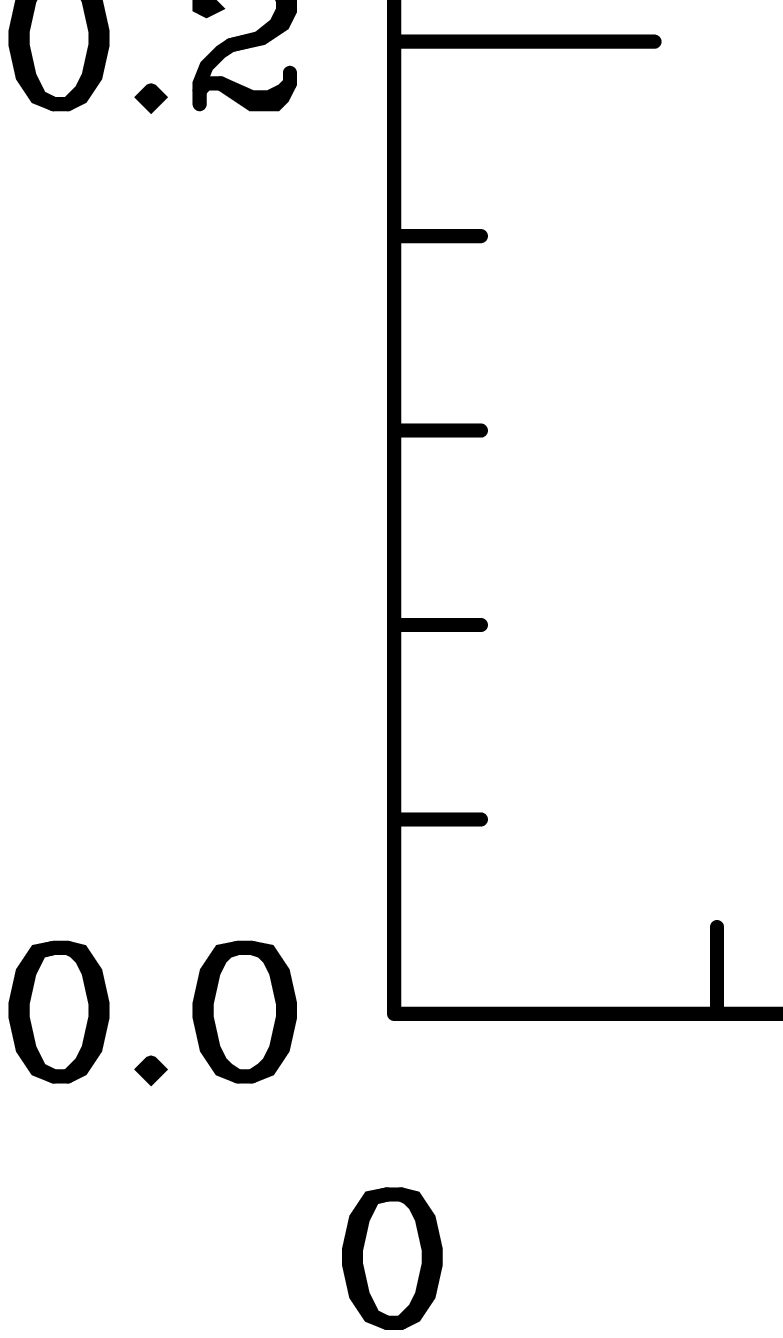}}
}
\caption{\label{fig:mimic}
The increments of a finite random walk with persistence (left)
may mimic certain aspects of the experimental data (right),
but with notable differences.  The random process has $a=0.8894$,
so an exponential decay time $\tau=4.00$.  The experiment is a
typical capacitance noise time series of HOSE, with a measured
$1/e$ crossing of $5.7$.  (a) and (b)
show the best-fit lines to the first 100 points (excluding
zero and the lowest frequency) of the power spectrum; both
give slopes $\approx-1.0$, but the random data level off
noticeably at low frequencies, as would be expected of white
noise.  Autocorrelation curves (c) and (d)
show fits to exponential (dotted line)
and shifted power-law \eqref{eqn:beta0} (solid) decays.  For
the random noise, the two fits fall on top of one another,
but for the experimental data, a power law fits
better than exponential decay.
}
\end{figure}

First, consider the increments of a discrete random walk with
persistence.
Let the increment at time $j\Delta t$, where $\Delta t$ is the
time step, be $x_j$, drawn from $\{+1,-1\}$.  Then $x_{j+1}=x_j$
with probability $a$ and $x_{j+1}=-x_j$ with probability $1-a$;
one recovers the usual discrete binary random walk for $a=1/2$.
Since we think of this process as approximating a continuous one,
and there is no natural way to take the limit $\Delta t\rightarrow0$
for anticorrelated increments, we restrict $1/2 \le a \le 1$.  
For convenience, we set $\Delta t=1$.
A simple inductive argument shows that
\begin{equation}
\label{eqn:expdecay}
\langle x_0 x_n\rangle=(2a-1)^n=\exp(-n/\tau)\rlap{\quad,}
\end{equation}
where the correlation time $\tau=-1/\ln(2a-1)$.
For times much larger than $\tau$, this Markov process looks like an ordinary
binary random walk with a rescaled time, and by the usual arguments
\cite{reichl}, the power spectrum approaches white noise, {\it i.e.}, it is
independent of frequency. However, for a finite run, the power spectrum may
mimic correlated (pink) noise even, surprisingly, for a $\tau$ as short as $4$
in a run as long as $4096$, as in Figure \ref{fig:mimic}a. However, the random
noise levels off noticeably at low frequencies, while the experimental data
(Figure \ref{fig:mimic}b) appear to follow a $1/f^\alpha$ power law to the
lowest frequencies \footnote{Indeed, our Figure \ref{fig:mimic}a resembles
Figure 6b of Reference \cite{Rangarajan00}. That process also has no true
long-time correlations.}. This supports the presence of correlations at all time
scales. The shortness of the low-frequency plateau in Figure \ref{fig:mimic}a
is misleading. To see more of the flat part of the spectrum, finer frequency
resolution is necessary. Taking larger windows, we can (at least for a run
longer than 4096) extend the graph many decades to the left and verify that the
spectrum remains flat (white), but at the cost of greater scatter. Parts (c)
and (d) of the figure show autocorrelation for random noise and experimental
data with fits to exponential decay (dotted) and the shifted power law
\eqref{eqn:beta0} (solid). The two fits fall on top of one another for the
process satisfying \eqref{eqn:expdecay}. That exponential decay does not
approximate the experimental data as well as the power law corroborates the
hypothesis of longer-than-short-time correlations.

Mandelbrot and van Ness \cite{Mandelbrot68} introduce the
notion of fractional Brownian motion with correlations between
increments separated by arbitrary time differences and with a
$1/f^\alpha$ power spectrum.  Rangarajan and Ding \cite{Rangarajan00}
describe a particularly simple way of generating a time series
of increments with such properties: start with a Gaussian-distributed
uncorrelated time series $\{x_j\}$, Fourier-transform, multiply by
$f^{-\alpha/2}$, and Fourier-transform back.  The resulting
process has a Hurst exponent given by
\begin{equation}
\label{eqn:rangahurst}
H=(1+\alpha)/2 \rlap{\quad.}
\end{equation}
Determination of exponents $\alpha$ and $H$ is subject to the usual numerical
vicissitudes, but Rangarajan and Ding argue that true long-ranged
processes should satisfy \eqref{eqn:rangahurst}
at least approximately. 

\begin{figure}[b]
\centerline{%
\hbox{\epsfxsize\hsize\epsfbox{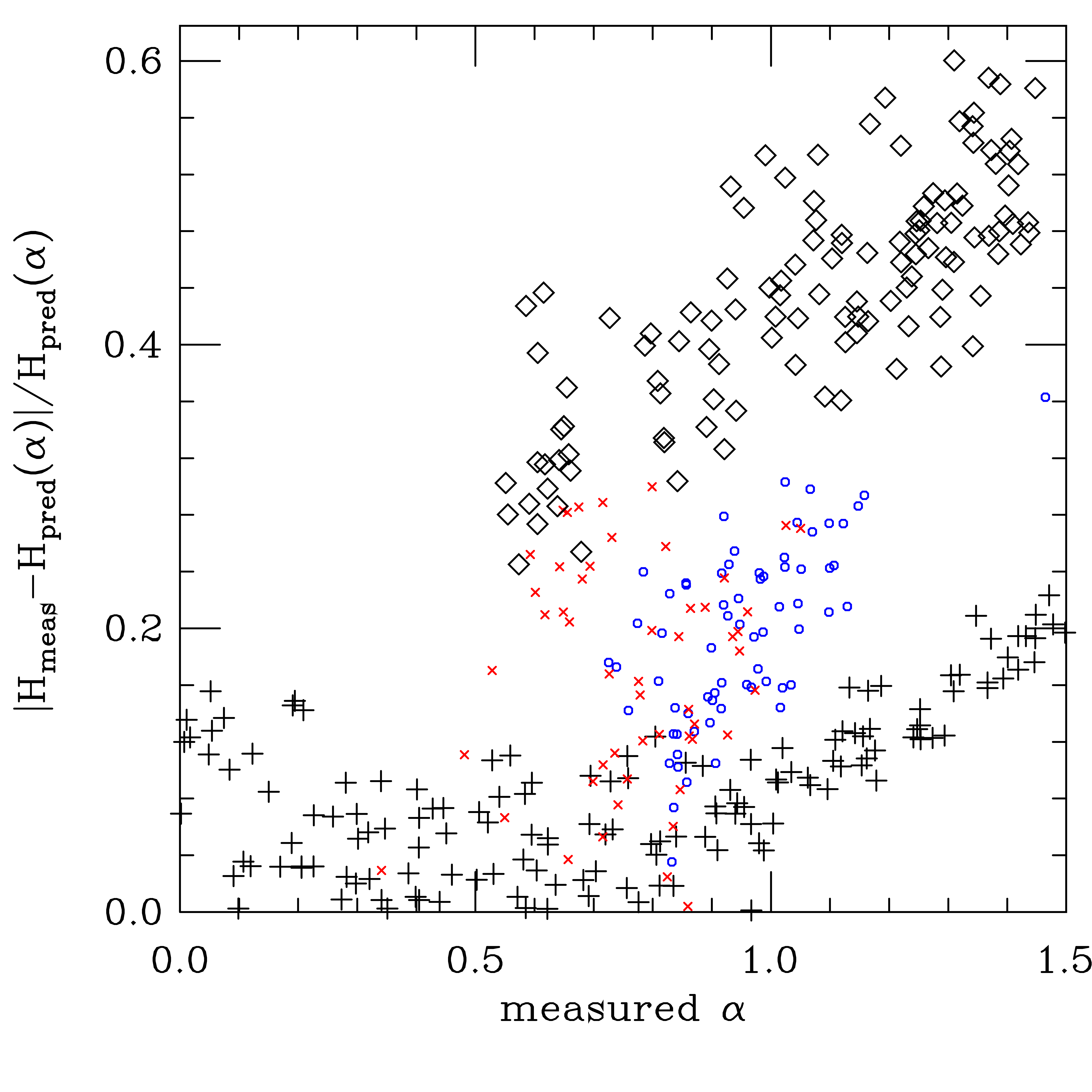}}
}
\caption{\label{fig:sandwich}
(Color on-line.)
Fractional discrepancies between $H_{\rm pred}$ given by
\eqref{eqn:rangahurst} and
measured Hurst exponent as functions of measured spectral
exponent $\alpha$.  Near the bottom, plotted with large $+$
symbols, are artificially-generated data with known
long-time correlations.
At top are generated data (large $\diamond$) from
random-walk increments with persistence times ranging from
$2$ (smaller values of $\alpha$) to $7$ (larger values).  In
the middle are experimental results for HOSE (blue $\circ$) and
SKOV (red $\times$).  Most of the experimental data look more like
the correlated data than the uncorrelated, but a few
overlap with uncorrelated noise; all of these are SKOV.
}
\end{figure}

Figure \ref{fig:sandwich} plots fractional discrepancies between
\eqref{eqn:rangahurst} and measured Hurst exponents as functions of
measured spectral exponents $\alpha$.  At the bottom
are plotted artificially-generated long-time-correlated data
following the prescription of Rangarajan and Ding (plotting symbols $+$); 
the measured exponents $\alpha$ are
always close to the known values, so the measurement errors
occur in estimating $H$.  We note a systematic trend toward
larger errors away from $\alpha\approx0.5$, but generally the
errors stay small.  At the top of the graph (plotting symbols $\diamond$)
are artificially-generated random walk increments with persistence
times ranging from $2$ at the left to $7$ at the right.  Measured
values of $\alpha$ follow the same prescription as used above,
although as noted earlier (Figure \ref{fig:mimic}), the fits 
fail for low frequencies; indeed, every $\alpha$ should
be zero.  Hurst estimates range from 0.45 to
0.67; the true value in every case should be $1/2$.  As discussed
by Rangarajan and Ding, the discrepancies between measured Hurst and
Hurst estimated from measured $\alpha$
are large.  In the middle and at the bottom are plotted
our experimental data (HOSE $\circ$, SKOV $\times$).  Agreement
between the exponents $H$ and $\alpha$ is generally not as
good as for the long-range-correlated processes but not so poor
as for the short-time-correlated random walk.  On average,
the experimental points lie closer to the former than to the latter.
We interpret this result as supporting the existence of
correlations on, at the very least, many different time scales.
A model of cell motion will need to explain both the
short-time and long-time correlations we have observed.

\section{APPLICATIONS}
We have demonstrated that electrical-noise measurements on
human ovarian surface epithelial cells
can distinguish cancerous and non-cancerous cultures.  This
is not intended as a diagnostic tool; for one thing,
it is easier to distinguish them under a microscope.  We find it
is also possible to distinguish HOSE from SKOV based
purely on {\it average\/} electrical resistance or
capacitance.  Our main focus has rather been on developing
statistical tools with which to test more sophisticated
statistical-mechanical models
and in developing a database of characteristics of many
different cell types, for which a single measurement ({\it e.g.},
average electrical resistance) will surely be inadequate.

One characteristic of malignant cancerous cells is their ability to invade
tissue in disregard of clues from their neighbors \cite{hanahan}; our
observation of shorter correlation times in cancerous cultures is consistent
with the picture of cancerous cells moving in a less regulated manner. Now
that it has been established that different cell types generate distinguishable
noise patterns, future research in this area will focus on the development of
realistic models of cellular motility for healthy and malignant cells.

\begin{acknowledgments}
We acknowledge the participation of Hiep Q.\ Le.
This work was supported in part by a grant from
the Florida Space Research Institute.
We would also like to thank Dr.\ Samuel Mok at 
Harvard Medical School for providing us with the cell lines used in this study.
Finally, we thank Heather Harper for helpful comments on the manuscript.
\end{acknowledgments}

\appendix
\section{Comparing discriminators}
\label{app:errors}
We claim no originality to the following elementary application
of statistics but could not find a textbook discussion of quite this point.
Given two distributions, $A$ and $B$ (for instance, the kurtoses
of HOSE data sets and those of SKOV), assumed to be Gaussian and
characterized by means $\mu_A<\mu_B$ and
standard deviations $\sigma_A$, $\sigma_B$, there are several
choices of where to place a dividing point $x_0$ so as to identify
all $x<x_0$ as belonging to population $1$ and all $x>x_0$ to
population $2$.  One natural choice is to pick $x_0$ so that
the expected rates of correct identification of the two populations
will be the same, {\it i.e.}, that $x_0-\mu_A$ should be the
same multiple of $\sigma_A$ as $\mu_B-x_0$ is of $\sigma_B$,
or
\begin{equation}
\label{eqn:x00}
x_0 = {{\mu_A\sigma_B + \mu_B\sigma_A}\over{\sigma_A+\sigma_B}}\rlap{\quad.}
\end{equation}
Any other choice will decrease the expected rate of incorrect identification
of one population at the cost of increasing the other.
A second plausible choice is to seek to maximize the sum of the
expected correct identification rates,
\begin{equation}
\label{eqn:errs}
\begin{aligned}
{\rm C}_A &= {1\over2}+{1\over2}{\rm erf}\left({{x_0-\mu_A}\over{\sigma_A\sqrt2}}\right)\\[0.5\baselineskip]
{\rm C}_B &= {1\over2}+{1\over2}{\rm erf}\left({{\mu_B-x_0}\over{\sigma_B\sqrt2}}\right)
\rlap{\quad;}
\end{aligned}
\end{equation}
it is easy to show that the separatrix $x_0$ is then
\begin{equation}
\label{eqn:x01}
\textstyle
x_0\!=\!{{\mu_B\sigma_A^2 - \sigma_B\left(\mu_A\sigma_B\pm 
\sigma_A\sqrt{(\mu_A-\mu_B)^2+2(\sigma_A^2-\sigma_B^2)\ln(\sigma_A/\sigma_B)}
\,\right)}
\over{\sigma_A^2-\sigma_B^2}}\rlap{\quad.}
\end{equation}
(One root maximizes ${\rm C}_A+{\rm C}_B$.  Note that \eqref{eqn:x01}
reduces to $(\mu_A+\mu_B)/2$ when $\sigma_A=\sigma_B$.)  A third natural choice,
maximizing the product ${\rm C}_A\,{\rm C}_B$, requires numerical
solution.  Of course, a more complicated risk function could apply,
for instance in medical diagnosis, where a false negative is much
worse than a false positive.

To compare the predictive values of three of the statistical measures
developed in the text, $1/e$ crossing from Table \ref{tab:auto},
kurtosis from Table \ref{tab:kurtosis}, and the Kolmogorov-Smirnov
test of Table \ref{tab:KS},
we apply the simplest separatrix, \eqref{eqn:x00}
to the means and standard deviations estimated for the first two. (This
choice is motivated by the similar correct-identification percentages
for HOSE and SKOV in Table \ref{tab:KS}, but as an alternative to
\eqref{eqn:errs}, using
the actual data sets gives comparable answers).  Then the expected
correct-identification
rate \eqref{eqn:errs} for $1/e$ as a discriminant is 62\%
and that for kurtosis 67\%.  These rates are both lower than
the 79\% (Table \ref{tab:KS}) for the Kolmogorov-Smirnov test
applied to the noise distribution, undermining the idea that
the deviation of this distribution from normal form is strictly
a proxy for correlation time.

\bibliography{distinguishing}
\end{document}